\documentclass[usenatbib]{mn2e}

\usepackage{graphicx}
\usepackage{amssymb}
\newcommand{\text}[1]{\quad\mbox{#1}\quad}

\newcommand{\oder}[2]{\frac{d #1}{d #2}}

\newcommand{\fracb}[2]{\left(\frac{#1}{#2}\right)}
\newcommand{\Ep}{{\cal E}_{\rm ph}}
\newcommand{\Ee}{{\cal E}_{\rm e}}
\newcommand{\mean}[1]{\langle{#1}\rangle}

\begin{document}

\title[Gamma-ray emission from the Crab Nebula]{On the origin of variable gamma-ray emission from the Crab Nebula}

\author[Komissarov \& Lyutikov]{S. S. Komissarov,$^{1}$\thanks{E-mail: serguei@maths.leeds.ac.uk (SSK)} 
M. Lyutikov$^{1}$\thanks{E-mail: lyutikov@purdue.edu}\\ 
$^{1}$Department of Applied Mathematics, The University of Leeds, Leeds, LS2 9GT \\ 
$^{2}$Department of Physics, Purdue University, 525 Northwestern Avenue, 
West Lafayette, IN 47907-2036, USA}

\date{Received/Accepted}
\maketitle

\begin{abstract}
The oblique geometry of pulsar wind termination shock ensures that 
the Doppler beaming has a strong impact on the shock emission. 
We illustrate this using the recent relativistic MHD simulations of the Crab 
Nebula and analysis of oblique shocks. 
We also show that the observed size, shape, 
and distance from the Crab pulsar of the Crab Nebula inner knot are consistent 
with its interpretation as a Doppler-boosted emission from the termination 
shock. If the electrons responsible for the synchrotron gamma-rays are accelerated 
only at the termination shock then their short life-time ensures that 
these gamma-rays originate close to the shock and are also strongly 
effected by the Doppler beaming.  As the result,
bulk of the observed synchrotron gamma-rays of the Crab 
Nebula around 100~MeV may come from its inner knot. 
This hypothesis is consistent with the observed optical flux of the inner knot,
provided its optical-gamma spectral index is the same as the injection 
spectral index found in the Kennel \& Coroniti model of the nebula spectrum. 
The observed variability of synchrotron gamma-ray emission on the time scale of 
wisp production can be caused by the instability of the termination shock discovered in 
recent numerical simulations.  Given the small size of the knot, it is also possible that 
the September 2010 gamma-ray flare of the Crab Nebula also came from the knot, though the 
actual mechanism remains unclear. The model predicts correlation of the temporal 
variability of the synchrotron gamma-ray flux in the Fermi and AGILE windows 
with the variability of the un-pulsed optical flux from within $1\arcsec$ of the 
Crab pulsar.                
\end{abstract}

\begin{keywords}
ISM: supernova remnants -- MHD -- shock waves -- gamma-rays: theory --
radiation mechanisms: non-thermal -- relativity -- pulsars: individual: Crab
\end{keywords}

\section{Introduction}
\label{sec:intr}

The Crab Nebula has been a source of intriguing discoveries and served 
as a test bed of astrophysics for decades. This is one of the best studied 
objects beyond the Solar system. It has been observed at all wavelengths, from 
radio to very high energy gamma-rays. Its non-thermal emission below 
$\Ep^{\rm b}\simeq500\,$MeV is a synchrotron emission of relativistic electrons 
in the nebula magnetic field and above $\Ep^{\rm b}$ it is the inverse 
Compton emission of the same electrons. The emitting electrons are accelerated 
up to PeV energies, indicating that the acceleration mechanism is very 
efficient. The source of energy is the ultra-relativistic magnetic wind 
from the pulsar \citep{RG74,KC84a}, but the actual mechanism of particle acceleration 
is still a mystery. 
The main candidates are the diffusive shock acceleration at the wind termination shock, 
the second-order Fermi acceleration in the turbulent plasma of the nebula, 
including secondary shocks, and the magnetic reconnection events. 

Compared to the highly filamentary thermal emission, the non-thermal 
emission is relatively featureless. Yet, it was discovered already 
in 1920 that fine and dynamic ``wisps'' are somehow  produced in the center 
of the nebula \citep{l21,s69}. Later, the X-ray observations discovered the 
famous jet-torus structure in the inner nebula \citep{W00,H02}, and the 
high resolution optical observations with Hubble Space Telescope revealed
fine sub-arcsecond structure of the non-thermal emission, including few 
optical knots \citep{H95,H02}.    

The synchrotron life-time of electrons emitting in optics is comparable to the 
dynamical time-scale of the nebula, and this makes it difficult to spot the 
exact locations of the particle acceleration cites. 
In gamma-rays below $\Ep^{\rm b}$, where the life-time 
becomes short, the angular resolution of the telescopes is insufficient to see 
the nebula structure. However, the observations indicated variability of the  
gamma-ray emission in the 1-150~MeV range \citep{M95,J96} on the time scale 
around one year. \citet{J96} proposed that this emission could originate from 
the variable optical features seen with HST in the polar regions of the 
inner nebula, in particular the so-called ``anvil''. Variability on a similar 
time-scale has been recently discovered in the X-ray emission \citep{WC10}.        

In September 2010,  AGILE collaboration 
reported a three-fold increase of the gamma-ray flux ($>$100~MeV) from the direction 
of the Crab Nebula \citep{T11}, which was immediately confirmed by Fermi LAT 
collaboration, who reported a six-fold increase of the flux \citep{A11}. 
The flare continued for four days, September 18-22, 
after which the gamma-ray flux returned to the pre-flare level. 
Fermi also reported that the pulsed emission of the Crab pulsar remained unchanged 
during the flare, suggesting that the flare originated in the Nebula.  
Jodrell Bank radio timing observations of the Crab pulsar showed no glitch 
during the flare, supporting this conclusion (ATel\#: 2889). 
At the same time, INTEGRAL reported no detection of the flare during Sep 19 in 
the 20-400~keV window (ATel\#: 2856) and Swift/BAT did not see any significant 
variability during the gamma-ray flare in the  14-150~keV range (ATel\#: 2893). 
Swift also reported no evidence for active AGN near the Crab, suggesting that 
the Crab itself is responsible for the flare (ATel\#: 2868). 
ARGO-YBJ collaboration reported a significant enhancement of the very high energy 
emission, around 1~TeV,  from the Crab nebula during the AGILE-Fermi 
flare (ATel\#: 2921). However, this has not been confirmed 
by VERITAS and MAGIC collaborations (ATel\#: 2967,2968). 
This discovery seems to have given credit to another event, detected in February 2009, 
which lasted for approximately 14 days, during which the gamma-ray flux increased 
by a factor of three or four \citep{T11,A11}. On the SED plots the flares appear as an 
extension of the synchrotron component further out towards the higher energies, up to 
1~GeV for the September flare and a bit less dramatic for the February flare.

The short duration of these flares suggests that their source is rather compact. 
Unfortunately,   
no high angular resolution observations of the nebula were carried out 
during the flares. The Crab Nebula images from Chandra and HST, obtained after the 
September flare, have not revealed 
anything especially unusual (ATel\#: 2882, ATel\#: 2903). They do show a change in 
the structure of the nebula wisps compared to previous observations, 
carried out years earlier. However, the large length 
scale of these wisps shows that they can hardly be a source of the flares.   
The Chandra images also show a significant change in the 
position of one of the jet knots, which apparently had moved about $3\arcsec$ 
towards the pulsar. This may be more significant as this feature 
is more compact.        

Even more potent source of the flares could be the mysterious ``inner knot'', 
discovered only $0\farcs65$ away from the Crab pulsar along the jet direction.
This knot, named as ``knot 1'' in \citet{H95}, is the brightest and most compact 
persistent feature of the Crab Nebula. 
It is seen at more or less the same location in many observations,  
both with space and ground based telescopes with adaptive optics, 
which followed its discovery. 
It is extended and elongated, with the main axis perpendicular to the jet.
Its length and width are $\psi_\perp \simeq 0\farcs50$ and
$\psi_\parallel\simeq 0\farcs16$ respectively \citep{H95}.
A number of recent optical and infra-red observations of the inner Crab Nebula
focused on the knot variability. \citet{M05} reported no variability on 
the timescales from 1 kilosecond to 48hr. 
\citet{T09} compared the measurements separated by two and half months and
found no significant difference too. On the other hand, 
\citet{S03} analyzed the HST archive data and discovered flux variations
on the level of 50$\%$ over the period of 6 years\footnote{For some reason the
actual observational data have not been presented in this paper.}. 
They also reported possible
random displacements of the knot on the scale of $0\farcs1$. More recently,
\citet{SS09} reported twice as higher flux from the inner knot in 2003 compared to
the earlier observations in 2000.  Thus, we tentatively conclude that the inner knot
does show significant variations of its flux, and possibly location, on the 
time-scale comparable to that of the gamma-rays variability reported by 
\citet{J96}.  These data suggest to consider the possibility that the inner 
knot may be a strong source of gamma-rays, both during and between the flares.  

The synthetic maps of synchrotron emission from the Crab Nebula, based on 
numerical relativistic MHD simulations, reveal
a bright compact feature strikingly similar in its 
appearance and location to the inner knot (see Figure~\ref{fig:sim}). 
In \citet{KL04} this feature was identified with the Doppler-boosted emission 
from the high-velocity flow downstream 
of the oblique termination shock of the pulsar wind.  
The more recent simulations of the Crab Nebula, which had a significantly 
higher resolution, discovered 
strong variability of the termination shock, involving dramatic 
changes of its shape and inclination \citep{C09}. This discovery 
suggests that the gamma-ray variability may be related to the changes in 
the Doppler beaming of the post-shock emission, associated with this structural 
variability of the termination shock. 
\citet{V08} have already addressed the
issue of variability of both the synchrotron and inverse Compton emission of 
the Crab Nebula in their computer simulations. They reported the strongest 
variability in the part of the gamma-ray window where the synchrotron emission is still dominant 
over the inverse Compton emission, and on the time-scale comparable 
to that found by \citep{J96}. However, they could not identify the source 
of variability with any particular feature in their numerical solutions.

\begin{figure}
\includegraphics[width=80mm,angle=0]{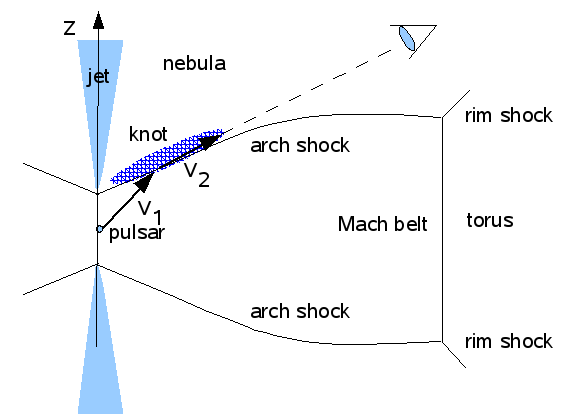}
\caption{Interpretation of the inner knot as the Doppler-boosted 
emission from the high velocity flow located downstream of the 
oblique termination shock. The termination shock is shown by the solid 
line. The dashed line is the line of sight.  
}
\label{fig:t-shock}
\end{figure}

Here we present new arguments in favour of the interpretation of the inner 
knot as a Doppler-boosted shock emission, and investigate whether the inner knot 
can be a strong source of variable gamma-ray emission.  
In Section~\ref{sec:origin} we analyse the observed shape and location of the 
inner knot and show that they are consistent with this interpretation.
In Section~\ref{sec:gamma} we argue that at least a significant 
fraction of the synchrotron gamma-ray emission of the Crab Nebula originates 
from the inner knot. The key factors are the short cooling time of electrons and 
the strong Doppler beaming of the emission originated in the vicinity of the 
termination shock.  In Section~\ref{sec:variability} we discuss the possible 
connection between the observed variability of the gamma-ray emission and the 
non-stationary shock dynamics discovered in numerical simulations, 
in particular the role of the variable Doppler beam orientation. 
Our conclusions are given in Section~\ref{sec:conclusions}.    
In Appendix~\ref{sec:obl} we present the analysis of oblique MHD shocks. Its 
results allow us to determine how high the Lorentz factor downstream of 
the Crab's termination shock can be and to confirm the results of numerical 
simulations.

\section{Origin of the inner knot} 
\label{sec:origin}

\begin{figure*}
\includegraphics[width=75mm,angle=0]{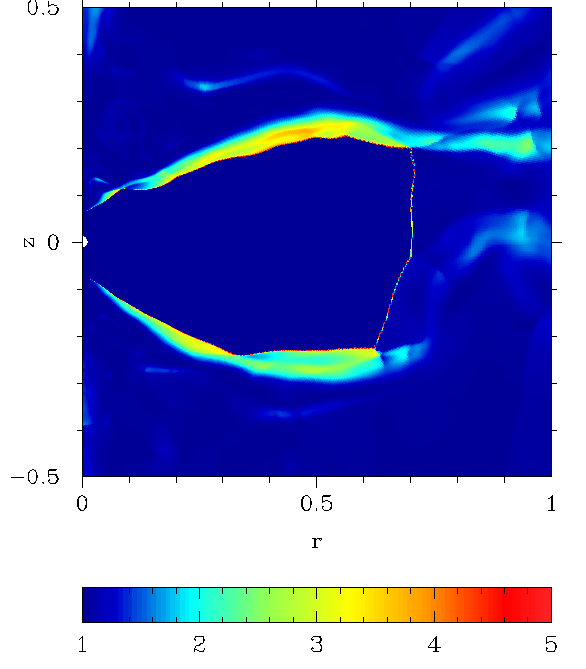}
\includegraphics[width=75mm,angle=0]{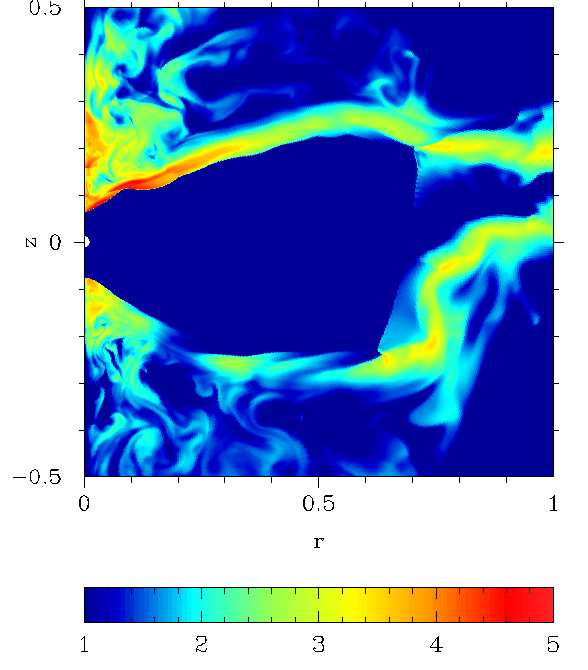}
\caption{
Termination shock in numerical simulations of \citet{C09}. 
The left panel shows the flow Lorentz factor. In this plot the wind zone 
has been cut off (The thin line at $r\sim 0.7$ is an artifact of this procedure, which 
conveniently indicates the location of the Mach belt). 
The right panel shows the observed synchrotron emissivity 
in the optical range, $\log_{10} j_{\nu,\rm ob}$, in the plane which includes 
the line of sight and the symmetry axis. 
The angle between the line of sight and the symmetry axis
is $60\degr$.}
\label{fig:sim}
\end{figure*}

\begin{figure*}
 \includegraphics[width=75mm,angle=0]{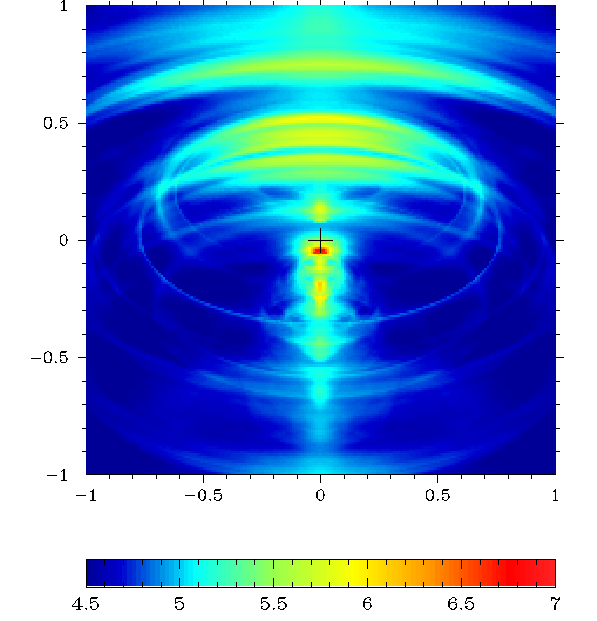}
 \includegraphics[width=75mm,angle=0]{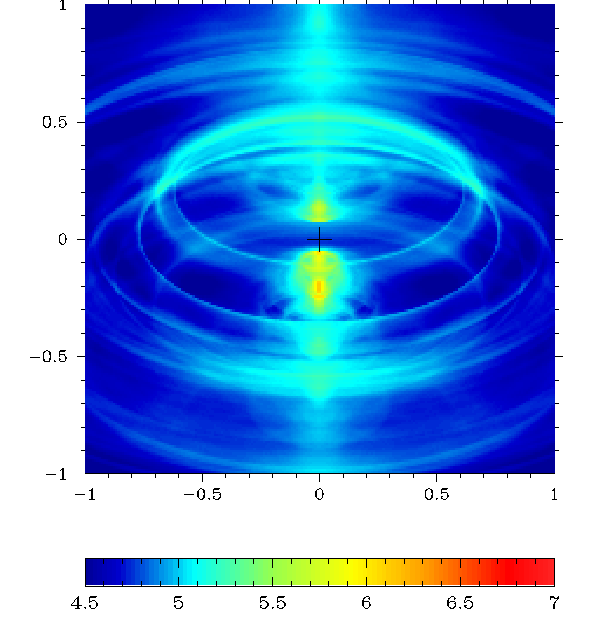}
\caption{
Synthetic images of the inner Crab nebula in optics.  
The  panel shows the proper synchrotron image, $\log_{10} I_\nu$, where $I_\nu$ 
is the intensity of radiation. The angle between the line of sight and the symmetry 
axis is $60\degr$. The right panel shows how the image would look if the 
the Doppler beaming was not taken into account. }
\label{fig:optics}
\end{figure*}

The first numerical simulations of the Crab Nebula by 
\citep{KL03,KL04,LDZ04,B05} had rather low numerical resolution 
and imposed reflectional symmetry in the equatorial plane. In these 
simulations the termination shock appeared as a more or less stationary 
complex structure, which in fact included several different shocks. 
These components, named as arch shocks, rim shocks, and the Mach belt 
in \citet{KL04}, are shown in Figure~\ref{fig:t-shock}. 
The latest simulations by \citet{C09} had much higher 
resolution and the computational domain included the whole range of 
the polar angle, $\theta\in(0,\pi)$. In these simulations, the structure of 
termination shock appeared highly distorted and dynamic, but as 
one can see in Figure~\ref{fig:sim}, these individual components were still 
identifiable.     

The right panel of Figure~\ref{fig:sim} shows the typical distribution of the 
synchrotron emissivity as measured in the pulsar/observer frame, found in the 
latest numerical simulations \footnote{  
The details of these numerical simulations and the radiation transfer 
computations are described in \citet{C09}. Here we only note that the radiation 
model assumes that relativistic electrons with the power law energy spectrum 
$N(\Ee)\propto \Ee^{2.2}$, terminated at $\Ee^{max}=1\,$PeV, are injected at the
termination shock, to be more precise at the arch shock and the Mach belt, and 
then they evolve subject to advection and synchrotron and adiabatic energy losses.}.  
One can see that it is strongly enhanced in the vicinity of the upper 
arch shock of the termination shock complex. There are two reasons for this 
enhancement. First, the proper magnetic field in this region is strongest. 
Several factors are responsible for this result. Although the  
the  magnetic field of dissipationless wind behaves as $B_\phi \propto \sin \theta/r$ 
\citep{M73,B99},  in the simulations this function was multiplied by 
$|1-2\theta/\pi|$, in order to account for the magnetic field dissipation 
in the striped wind zone. Thus, the wind magnetic field peaked at $\theta\simeq 57\degr$ 
instead of $\theta\simeq 90\degr$. More important, however, is  
the axial compression of the nebula by the magnetic hoop stress, which leads to the  
total pressure downstream of the termination shock to be significantly higher at small 
polar angles. 
As the result, the arch shock is pushed closer to the pulsar, leading to a stronger 
upstream and hence downstream magnetic field.    
Second, the emission from the upper arch shock is Doppler boosted. 
Indeed, as one can see in the left panel of Figure~\ref{fig:sim}, 
the Lorentz factor of the flow downstream of the arch shocks is quite high. 

The analysis of oblique relativistic MHD 
shocks given in Appendix~\ref{sec:obl} shows that in the case of 
ultra-relativistic cold upstream flow the downstream Lorentz factor is 
\begin{equation}
   \gamma_2  = \frac{1}{\sqrt{1-\chi^2}} \frac{1}{\sin\delta_1},
\label{gamma2}
\end{equation}
where $\delta_1$ is the angle between the upstream velocity and the shock 
plane and 
\begin{equation}
\chi=\frac{1+2\sigma_1 + \sqrt{16\sigma_1^2+16\sigma_1+1}}{6(1+\sigma_1)},
\label{chi}
\end{equation}
where $\sigma_1=B_1^2/4\pi\rho_1c^2$ is the magnetization parameter of the upstream 
flow. For $\sigma_1\gg 1$ this yields
\begin{equation}
   \gamma_2  \simeq \frac{\sigma_1^{1/2}}{\sin\delta_1},
\label{lim1}
\end{equation}
and for $\sigma_1\ll 1$
\begin{equation}
   \gamma_2  \simeq \frac{3}{2\sqrt{2}} (1+\frac{1}{2}\sigma_1)
   \frac{1}{\sin\delta_1}.
\label{lim2}
\end{equation}
In the simulations $\sigma_1$ varies with the polar angle between 0 and 0.05, with 
the volume averaged value $\mean{\sigma_1}\simeq 0.014$, and thus the latter limit 
applies. On can see that for small shock inclination angles the Lorentz factor 
can indeed be quite high. Even higher values are expected for high-sigma pulsar wind. 

In addition to having high Lorentz factor in the downstream flow, the upper arch shock is 
inclined at the angle of $\sim 60\degr$ to the polar axis, near the axis. 
Observations of the inner Crab Nebula 
suggest that the angle between the line of sight and 
the symmetry axis of the nebula is also close to $60\degr$ \citep{W00}. 
Thus, the upper arch shock is well aligned with the line of sight, resulting in 
strong Doppler-boosting of its emission. 
This is schematically illustrated in Figure~\ref{fig:t-shock}.   
The left panel of Figure~\ref{fig:optics} shows the synthetic optical synchrotron 
image of the inner part of the simulated PWN at the time corresponding to the age of 
the Crab Nebula. 
One can see prominent wisps and a bright knot located very close 
to the origin, where the projected image of the pulsar would appear if it was 
included in the emission model. 
In the simulations, there is no emission from the pulsar wind as the emitting 
electrons are injected at the termination shock only. Thus, all the fine features 
of the synthetic synchrotron images, including the inner knot, are produces 
inside the nebula.  

\begin{figure*}
\begin{center}
 \includegraphics[width=105mm,angle=0]{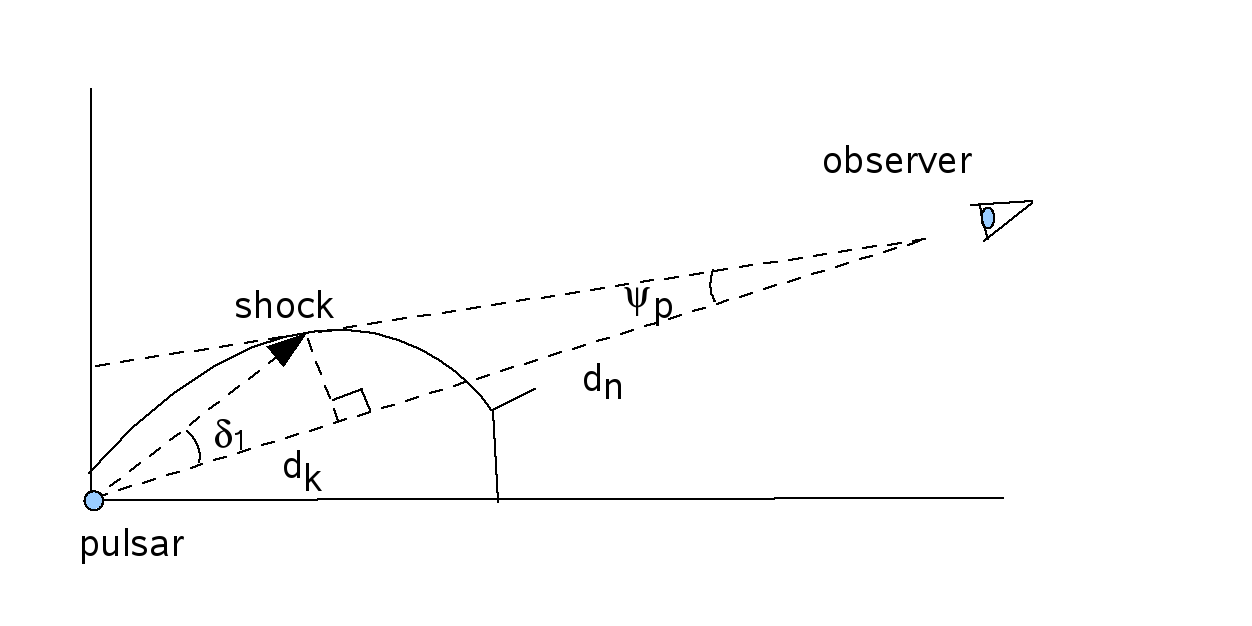}
 \includegraphics[width=55mm,angle=0]{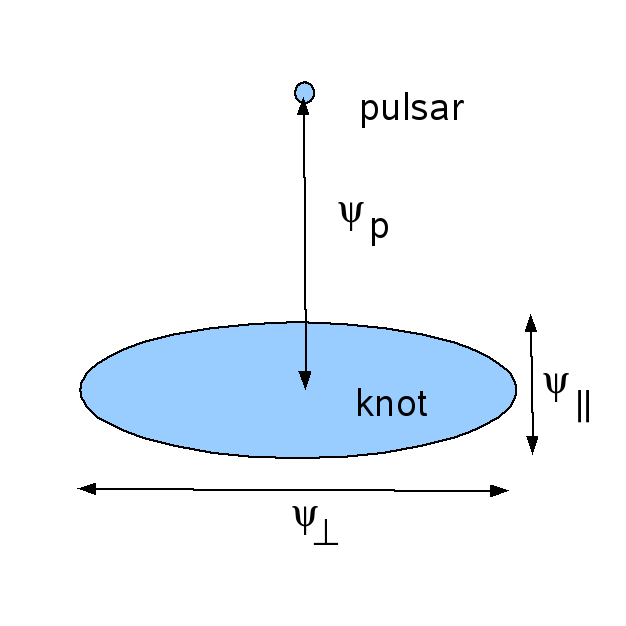}
\end{center}
\caption{
Basic geometric parameters of the inner knot. }
\label{fig:knot1}
\end{figure*}

It is interesting that both in the synthetic and the real optical images of the 
Crab Nebula the termination shock is not clearly identifiable\footnote{The Chandra X-ray 
image of the inner Crab Nebular is overall similar to the HST image \citep{H02}. 
There are however some noticeable differences. In particular, the X-ray image is much 
knottier and some of these X-ray knots arrange in a sort of ring, called the ``inner ring'', 
around the pulsar. This ring is often identified with the termination shock. The fact 
that the ring is relatively symmetric, in contrast with the optical image, is hard to 
explain in our model and indicates that some other factors have to be included.}. 
The inner wisps give away its size but that is about it. Several factors contribute 
to this effect. Firstly, the inner cavity filled with the pulsar wind is small 
compared to the size of the nebula, making the central brightness reduction rather 
weak. Secondly, the region of enhanced proper emissivity around the termination shock   
forms a geometrically thin distorted shell. Because of this, the observed emission 
is strongly enhanced in places where the line of sight is tangent 
to the shell surface. This leads to the appearance of several bright rings 
on synthetic maps where the Doppler beaming is not included (see the right panel 
of Figure~\ref{fig:optics}). The Doppler beaming leads to increased emissivity in the 
part of such a ring where its plasma flows towards the observer, and decreased emissivity 
where it flows away, turning the ring into an arc 
(see the left panel of Figure~\ref{fig:optics}). Because of the non-spherical shape 
of the termination shock, the jet base is located much closer to the pulsar 
compared to the Mach belt radius, making the wrong impression that the jet originates 
from the inside of the termination shock and suggesting that it is produced by the 
pulsar.    

Comparing the images presented in Figure~\ref{fig:optics} one can 
see how some features brighten up and others get dimmer because of the Doppler beaming.     
In particularly, the bright knot in the center of the left image is certainly  
Doppler-boosted.  \citet{KL04} proposed that this synthetic knot 
is a counterpart of the Crab's inner knot. Indeed, like the inner knot it is 
positioned on the jet-side of the nebula at the base of the jet (only in projection) 
and it is elongated in the direction perpendicular to the jet \citep{H95}.  
Other synthetic jet knots, which are seen 
in Figure~\ref{fig:optics}, have more or less the same brightness in both these images, 
indicating that the Doppler beaming is not that important. They indeed originate at 
the base of the polar jets.  The jet Lorentz factor is variable but on average it is 
rather low, $\gamma_j\sim 1.5$. Combined with the large viewing angle, $60\degr$, 
this explain why the Doppler effect is rather weak.  Phenomenologically, these 
jet knots are created via unsteady inhomogeneous axial magnetic pinch, which is 
responsible for the jet formation. The flow towards the polar axis, which feeds the jet, 
is highly inhomogeneous with patches of strong and weak magnetic field, resulting in strong 
spatial variations of the magnetic hoop stress and highly variable jet 
dynamics. The non-linear sausage-mode of the magnetic pinch instability could be 
another contributor to the jet variability. 

The fact that the inner knot of the Crab Nebula is so compact and that it appears  
at the base of the Crab jet are more suggestive of some jet feature rather than 
the extended termination shock. However, the knot shape and size are nicely explained in 
the shock model.  Denote as $\psi_p$ the observed angular distance of the knot from the pulsar, 
as $d_n$ the linear distance to the nebula, and as $d_k$ the linear 
distance between the  pulsar and the point on the shock surface 
where the line of sight is tangent to the shock. The angle between the shock 
surface and the upstream velocity vector denote as $\delta_1$ (see Fig.\ref{fig:knot1}). 
Then in the small angle approximation
\begin{equation}
    \delta_1 = \psi_p (d_n/d_k). 
\label{delta1}
\end{equation}
Using Eq.\ref{gamma2} we can now find the Lorentz factor of the 
postshock flow 
\begin{equation}
     {\gamma_2}=\frac{f_\sigma}{\delta_1}
\label{gamma2a}
\end{equation}
and the beaming angle
\begin{equation}
     \phi_D=\frac{1}{\gamma_2}=\frac{\delta_1}{f_\sigma},
\label{b-angle}
\end{equation}
where $f_\sigma=(1-\chi(\sigma_1))^{-1/2}$. 
This allows us to find the transverse angular size of the knot  
\begin{equation}
   \psi_\perp = \frac{d_k \phi_D}{d_n} = \frac{1}{f_\sigma} \psi_p. 
\end{equation}
For $\sigma_1\le 1$, one has $f_\sigma \simeq 1$ and, thus,     
\begin{equation}
   \psi_\perp \simeq \psi_p,
\label{db1}
\end{equation}
which is in excellent agreement with the observations of the inner knot. 
The same argument shows that HST knot 2, which has similar size but is located 
much further out from the pulsar, cannot be the Doppler-boosted part of the termination 
shock, unless the pulsar wind magnetization $\sigma_1\gg 1$.

The other size of the inner knot, $\psi_\parallel$, is determined by the 
thickness of the post-shock plasma flow and the shock local curvature. 
The flow thickness at the distance $d_k$ from the pulsar 
can be estimated as $\delta_2 d_k$, where $\delta_2$ is the downstream angle between 
the velocity vector and the shock plane. 
Since the distance of the shock from the pulsar across the line of sight is 
$\delta_1 d_k$, this yields $\psi_\parallel = (\delta_2/\delta_1)\psi_p$. 
Using Eq.\ref{os13} to evaluate $\delta_2/\delta_1$ and Eq.\ref{db1} we then find 
that if $\psi_\parallel$ is fully determined by the thickness of the post-shock flow  
then 
\begin{equation}
   \psi_\parallel \simeq \frac{1}{3} \psi_\perp, 
\label{db2}
\end{equation}
in excellent agreement with the observations again. 

The shock curvature would lead to a finite linear width of the knot  
in projection on the plane of the sky even if the postshock flow was 
infinitely thin. This width can be estimated as $R_c(1-\cos(\phi_D/2))$, 
where $R_c$ is the local shock curvature radius.
For the small angles this yields 
\begin{equation}
   \psi_\parallel \simeq \frac{(R_c/d_k)}{8\gamma_2 f_\sigma} \psi_\perp.
\label{db3}
\end{equation}
Thus, unless $(R_c/d_k)>8\gamma_2 f_\sigma$, we have $\psi_\parallel<\psi_\perp$, in 
agreement with the observations. 
Numerical simulations show that normally $R_c/d_k<10$.   

The transverse angular size of the knot could have been used to infer the 
Lorentz factor of the post-shock flow if we knew the distance $d_k$ between 
the knot and the pulsar. Indeed, combining Eqs.~\ref{gamma2a} and 
\ref{delta1} one finds that  
$$
  \gamma_2 \simeq \frac{d_k}{d_n} \frac{1}{\psi_\perp}.  
$$
Since we only know that $d_k$ cannot exceed the radius of the termination shock, 
this equation only allows us to find the upper limit 
$$
  \gamma_2 < \frac{\psi_{\rm ts}}{\psi_\perp} \simeq 20,
$$
where $\psi_{\rm ts}\simeq 10\arcsec$ is the angular size of termination shock inferred from the 
observations \citep{H02}. 

If the inner knot is indeed a part of the termination shock then its spectrum 
can be used to infer the properties of the particle acceleration at the shock. 
Several groups have carried out optical and near-infrared observations 
of the knot in recent years. Unfortunately, their results do not quite agree. 
\citet{SS09} reported the optical spectral index $\alpha=1.3$, 
assuming that $I_\nu\propto \nu^{-\alpha}$.  
On the other hand, the data presented in their Fig.2 suggest a much flatter 
near-infrared spectrum, with $\alpha \sim 0.3$. It is difficult to see 
how the synchrotron mechanism can accommodate such a large variation of 
spectral index within only one decade of frequency. Perhaps, these measurements 
suffer from large systematic errors. The proximity of the knot to the 
pulsar could be one of the complications. According to the data obtained  
by \citet{M05} the near-infrared spectral index of the inner knot is $\alpha=0.78$.
Unfortunately, the accuracy of this measurement is not given, whereas 
for other features the error is given and it is about $\pm0.13$. The proximity 
to the pulsar suggests that for the inner knot the error is higher. Finally, 
\citet{T09} give $\alpha=0.63\pm0.02$ for the optical emission of the inner knot. 
The very small error indicates that this is the most accurate measurement to date. 
This result is in excellent agreement with the value of the injection spectral 
index inferred by \citet{KC84} via model fitting of the integral spectrum of 
the Crab Nebula.

Although, the MHD model in general, and the recent numerical MHD simulations in 
particular, have enjoyed a lot of success in explaining the 
properties of the Crab Nebula, as well as other PWN, it is by no means 
problems free. The so-called $\sigma$-problem is its main weakness. It is not 
clear how exactly the pulsar wind turns from being Poynting-dominated near 
the pulsar to kinetic-energy-dominated near the termination shock. A number of 
different ideas have been put forward but the issue is far from settled 
\citep{LK01,KS03,L03,L03a,A08}. 
An alternative model, where the flow remains Poynting-dominated
even inside the nebula, has been put forward recently \citep{Lt10}. 
As the result, it is not clear as to what model of the pulsar wind should be 
used in setting the inflow boundary condition in the MHD simulations of PWN. 
Moreover, so far only two-dimensional simulation have been carried out, which 
leaves unexplored the effects of non-axisymmetric instabilities on the nebula 
structure and dynamics \citep{B98}.   

How and where the emitting particles are accelerated is also debated. 
In the simulations, it was assumed that the synchrotron electrons   
are accelerated only at the arch shock and the Mach belt, but the acceleration can 
also occur at the rim shocks and in the turbulent interior of the nebula.

Given these uncertainties, it is not surprising that there are still 
some significant quantitative differences between the theory and the observations.     
For example, the observations reveal comparable isotropic luminosities of the inner 
knot and the brightest wisps \citep{H95}. In contrast, in the synthetic synchrotron 
images the knot is several times brighter. This is illustrated in  
Figure~\ref{fig:sim}, which shows the intensity of radiation using  
logarithmic scaling. 
One possible reason for this is the excessive axial compression of the nebula 
via the hoop stress of the azimuthal magnetic field, caused by the condition 
of axisymmetry. In fact, the synthetic optical images of the Crab Nebula show 
strong global enhancement of the surface brightness along the 
symmetry axis, which is not present in the images of the Crab Nebula. 
The observed large scale kink of the Crab jet indicates that some kind of 
kink-mode instability significantly reduces the degree of symmetry 
in the polar regions of the nebula. This is important as, in addition to 
the anisotropic power distribution in the pulsar wind, the overall geometry of the 
termination shock is also influenced by the pressure distribution inside the nebula. 
The enhanced pressure near the axis pushes the cusp of the arch shock further down 
towards the pulsar, making the termination shock less spherical compared to what it  
would be in the case of uniform pressure distribution inside the nebula. 
This leads to higher magnetic field and 
number density of emitting particles, and hence increased volume emissivity 
near the shock cusp. In addition, the shock geometry determines 
the velocity distribution and hence the Doppler beaming. 

In three dimensional simulations of the Crab Nebula, the almost perfect 
alignment of the arch shock with the line of sigh may no longer be found. 
Less squashed along the polar direction, the termination shock would have lower 
proper emissivity near the cusp region. The Doppler-boosting of the knot 
emission will be reduced as well, and not only because of the less perfect 
alignment of its Doppler beam with the line of sight.  More spherical shape of 
the termination shock would also lead to higher shock inclination angles and 
lower Lorentz factor of the downstream flow.

\section{Why the inner knot can be a strong source of gamma-ray emission}
\label{sec:gamma}

\begin{figure}
 \includegraphics[width=80mm,angle=0]{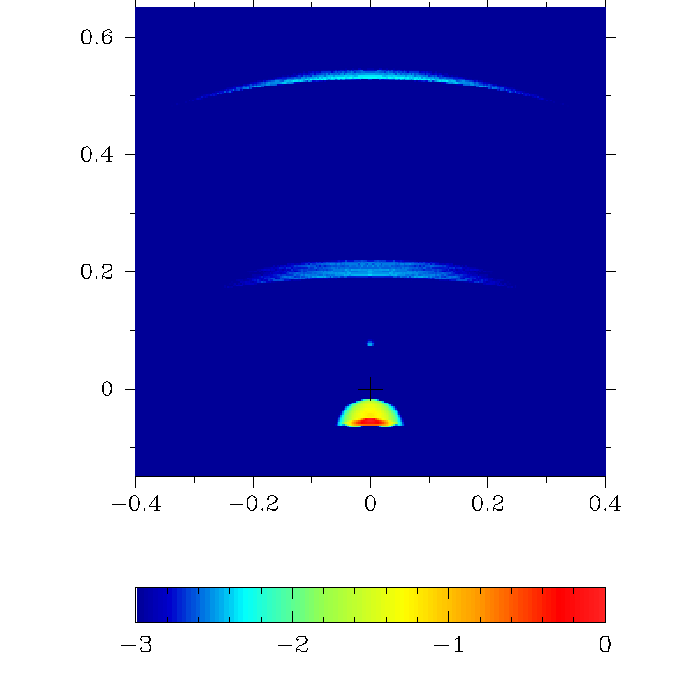}
\caption{Synthetic synchrotron map of the Crab Nebula at 100~MeV.
The image shows $\log_{10} I_\nu$ in arbitrary units. 
}
\label{fig:100MeV}
\end{figure}

Suppose that the termination shock is the main acceleration site of gamma-ray 
emitting electrons. The synchrotron cooling timescale is 
$$
   t_{\rm cool} \simeq 3.7 D^{1/2} \fracb{B}{10^3\mbox{G}}^{-3/2} 
   \fracb{{\cal E}_{\rm ph,ob}}{100\,\mbox{MeV}}^{-1/2} \mbox{days},  
$$    
where ${\cal E}_{\rm ph,ob}=D {\cal E}_{\rm ph}$ is the observed energy 
of photons emitted at the energy ${\cal E}_{\rm ph}$ in the fluid  
frame of the downstream plasma, 
$$
   D=\frac{1}{\gamma(1-\beta\cos\phi)} 
$$
is the Doppler factor, $B$ is the magnetic field strength as measured in 
the fluid frame, and $\phi$ is the angle between the line of sight 
and the velocity vector of the plasma bulk motion. 
Fitting of the Crab Nebula spectrum with the synchro-Compton 
model yields the typical $B\sim 100-200\,\mu$G \citep{HA04,A09}. However, 
individual bright features can have stronger magnetic field. In particular, 
\citet{H95} give the equipartition $B\simeq 2.5\times10^{-3}\mbox{G}$ for the inner 
knot. Moreover, the numerical simulations show that near the 
arch shock the magnetic field can be significantly higher, up to ten times, 
than the volume averaged\footnote{This is 
in contrast with the one-dimensional MHD model where the magnetic field increases 
with the distance from the termination shock \citep{KC84a}}. 
Thus, even for the Doppler factor as high as $D=10$, the cooling length 
scale of electrons emitting at $\Ep \sim 100\,$MeV is likely to be small compared  
to the termination shock radius, 
which is about $\simeq 10\arcsec$ or $\simeq 120\,$light days \citep{H02} in linear scale. 
Thus, the gamma-ray emitting region must be located very close to the shock.

Downstream of the Mach belt the magnetic field and the number density of the 
emitting particles is significantly lower compared to the arch shock, and for this 
reason the synchrotron emissivity is low as well (see Figure~\ref{fig:sim}). 
Thus, it is the wind plasma which has just passed through the arch shock 
which is likely to be the main emitter of 
the synchrotron gamma-rays. This plasma flows with relativistic speed and is subject 
to strong Doppler beaming (see Figure~\ref{fig:sim}). This results in boosting of the  
emission from the part of the flow where the velocity vector is close to line of sight, 
the inner knot region, and dimming of the emission from other parts, where the viewing angle 
exceeds $2/\gamma_2$. Thus, 
a significant fraction, if not most, of the observed synchrotron gamma-ray emission 
of the Crab Nebula may originate from its inner knot.    
In fact, quick inspection of our numerical solutions shows that at 100~MeV the 
inner knot is essentially the only feature in the sky (see Figure~\ref{fig:100MeV}). 
However, given the uncertainties of the numerical 
model, one cannot exclude a contribution from few brightest wisps.  

In order to test this idea against the 
observations one can compare the observed flux from the Crab's inner knot in optics 
with the observed flux from the whole of the Crab Nebula at 100~MeV. 
Given the small light crossing time of the knot compared to the synchrotron cooling 
time even at $100\,$MeV, its synchrotron electrons must still have the energy spectrum 
which is very close to the one produced by the shock acceleration mechanism.  
At $\nu=3.76\times10^{14}$Hz 
the de-reddened flux from the inner knot is 
$F_\nu\simeq1.6\times10^{-27}\mbox{erg}\,\mbox{s}^{-1}\mbox{cm}^{-2}\mbox{Hz}^{-1}$ 
\citep{T09}. At $100\,$MeV the observed flux is 
$F_\nu \simeq 1.7\times10^{-32}\mbox{erg}\,\mbox{s}^{-1}\mbox{cm}^{-2}\mbox{Hz}^{-1}$ 
\citep{A09}. The corresponding two point spectral index is $\alpha\simeq 0.64$. 
This is indeed the value of the injection spectral index inferred by \citet{KC84}, 
and measured in optics by \citet{T09}!    

There exists an upper limit on the energy of synchrotron photons, which is independent 
on the details of the acceleration mechanism \citep{Lt10}. If the accelerating electric field $E$ 
is a fraction $\eta \leq 1$ of the magnetic field then the rate of energy gain can be 
estimated as
\begin{equation}
\oder{\Ee}{t}=eEc= \eta e B c.
\end{equation}
The corresponding acceleration time scale $\tau_{\rm acc} = \Ee/(d\Ee/dt)=(\eta \omega_{\rm B})^{-1}$,
where $\omega_{\rm B}=ceB/\Ee$ is the relativistic Larmor frequency. 
The energy loss rate due to synchrotron emission 
\begin{equation}
\oder{\Ee}{t} = -c_2 B^2 \Ee^2,
\end{equation}
where $c_2=4e^4/9m^4c^7$ and we also assumed effective pitch angle scattering, grows 
with the electron energy. The balance of energy gains and losses yields the 
maximum energy, which can be reached by the accelerated electrons
\begin{equation}
{\cal E}_{\rm e}^{\rm max}=(\eta ec/c_2B)^{1/2}.
\end{equation}
The characteristic energy of the synchrotron photons emitted by the electron
of energy $\Ee^{\rm max}$ in the magnetic field of strength $B$,  
\begin{equation}
  {\cal E}_{\rm ph}^{\rm max} = c_1 B (\Ee^{\rm max})^2 = 
   \frac{27}{16\pi}\, \eta\, \frac{mhc^3}{e^2} = 236\,  \eta\, \mbox{MeV}\, ,
\end{equation}
where $c_1=3eh/4\pi m^3c^5$, does not depend on the magnetic field strength.  
This is the utmost upper limit, which may be impossible to reach in practice.
For example, \citet{J96} give an almost ten times
smaller value for $\Ep^{\rm max}$ for the shock acceleration.

In fact, the synchrotron component of the Crab Nebula spectrum becomes very steep 
above $10\,$MeV, and can be fitted with the function  
\begin{equation}
   F_\nu \propto \nu^{-\alpha} \exp(-h\nu/{\cal E}_{\rm ph}^{\rm c}), 
\label{cut-off}
\end{equation}
where the cut-off energy  
$\Ep^{\rm c}\simeq 100\,$MeV \citep{A09}. This is surprisingly close to 
our value of $\Ep^{\rm max}$. However, the above limit applies only 
in the frame of emitting plasma. If the plasma is moving with relativistic 
bulk speed relative to the observer then it has to be multiplied by the 
Doppler factor in order to obtained the corresponding observed photon energy. 
%
%
For $\gamma\gg 1$ the maximum value of 
the Doppler factor is 
$D_{\rm max} \simeq 2\gamma$, and thus even for the rather moderate 
postshock value of $\gamma_2\simeq 5$ the synchrotron cutoff energy can be 
increased by a factor of ten. This has been used to argue that the 
observed synchrotron emission of the Crab Nebula 
with ${\cal E}_{\rm ph} \gtrsim 100\,$MeV 
originates in relativistic flow  \citep{Lt10}.
Here, this argument can be refined to support the inner knot as the source of 
this gamma-ray emission because we are now almost certain that its emission 
is indeed Doppler-boosted.

It seems reasonable to expect the gyroradius of electrons accelerated at the termination 
shock to be below the radius of the termination shock.  An effective Doppler boosting 
may also require the gyroradius to be below the transverse size of the fast flow downstream of 
the arch-shock. Only in this case one can firmly conclude that the electrons are advected with 
the flow. For the electrons emitting synchrotron photons of the energy $\Ep$ in the 
comoving frame, the gyroradius radius is 
\begin{equation}
r_{\rm L} = \fracb{\Ep }{c_1 e^2 B^3}^{1/2}. 
\end{equation}
Our numerical simulations show that the magnetic field measured just downstream of the arch shock 
is significantly higher than the volume averaged one, up to about one order of magnitude. 
This indicates that $B=10^{-3}\,$G may well be typical for this region. Then the 
typical gyroradius radius of the electrons is   
\begin{equation}   
     r_{\rm L} \simeq \frac{1.6}{D^{1/2}}\fracb{{\cal E}_{\rm ph,ob}}{100\,\mbox{MeV}}^{1/2}
          \fracb{B}{10^{-3}\mbox{G}}^{-3/2}\, \mbox{light days}, 
\nonumber
\end{equation}
which is significantly less than the termination shock radius and even below  
the size $l_\parallel\simeq 2$~light days of the inner knot, 
which is an observational indicator of the thickness of the fast post-shock 
flow (see the discussion leading to Eq.\ref{db2}). 
Given this result, we conclude that the observed emission up to 
${\cal E}_{\rm ph,ob}=1\,$GeV can be the Doppler-boosted emission produced by the electrons 
accelerated at the termination shock.

\section{Nature of the gamma-ray variability}
\label{sec:variability}

\begin{figure*}
 \includegraphics[width=55mm,angle=0]{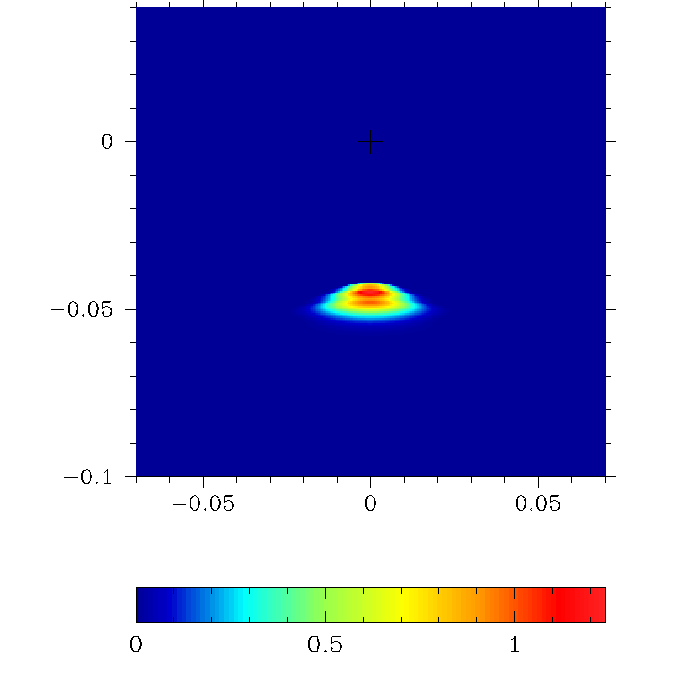}
 \includegraphics[width=55mm,angle=0]{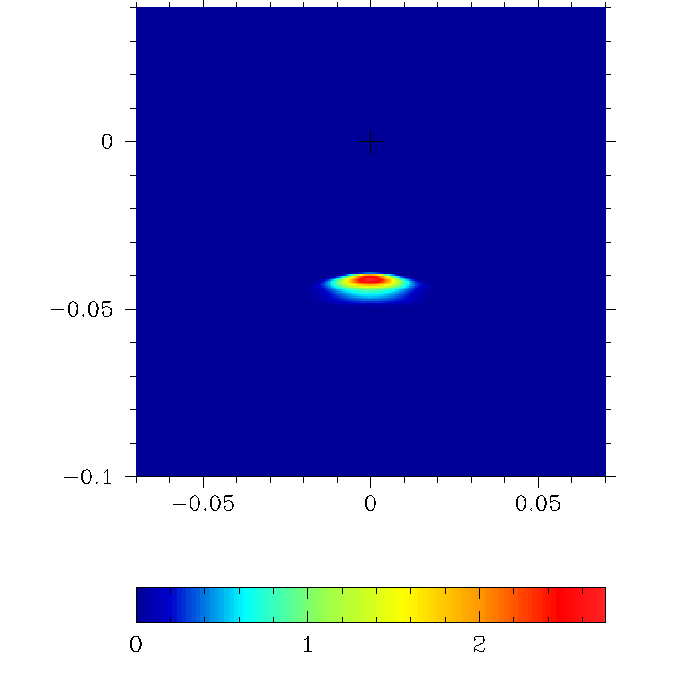}
 \includegraphics[width=55mm,angle=0]{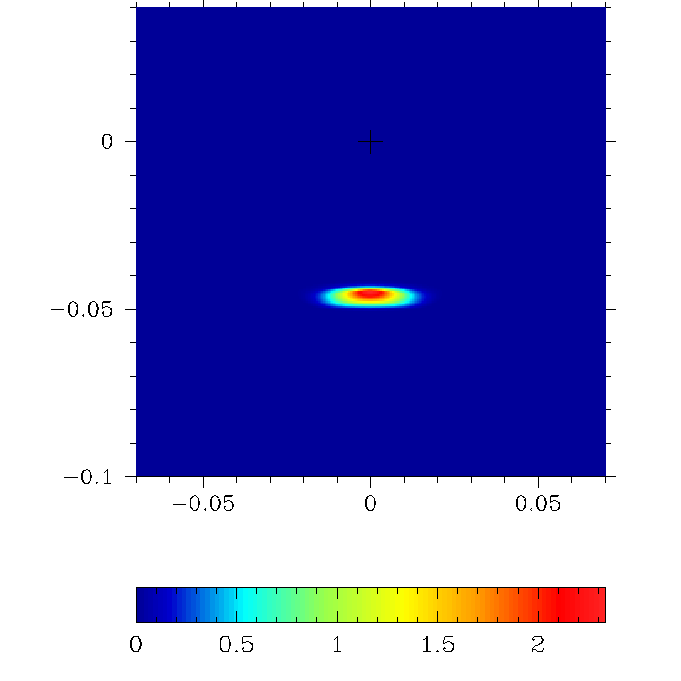}
 \includegraphics[width=55mm,angle=0]{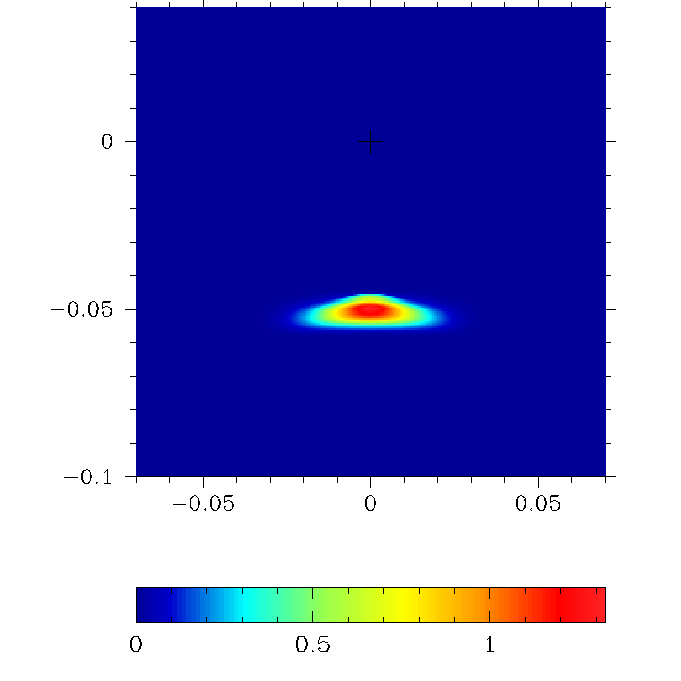}
 \includegraphics[width=55mm,angle=0]{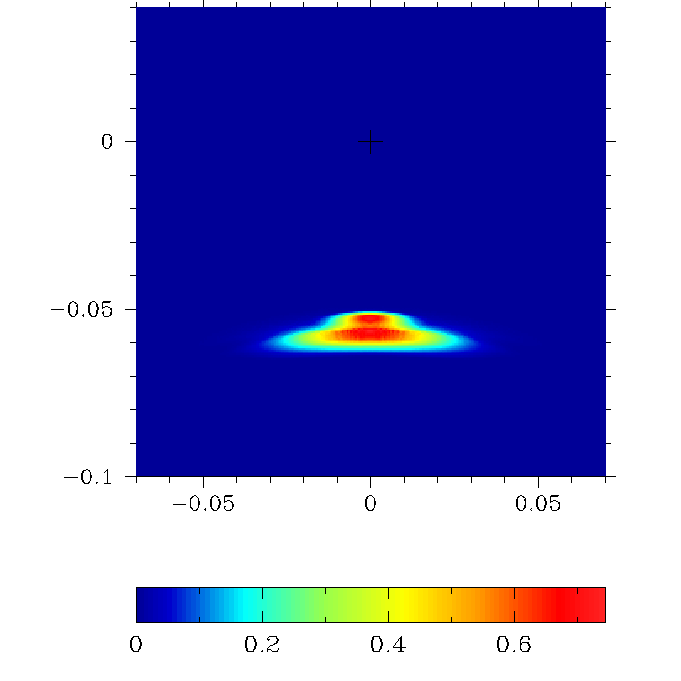}
 \includegraphics[width=55mm,angle=0]{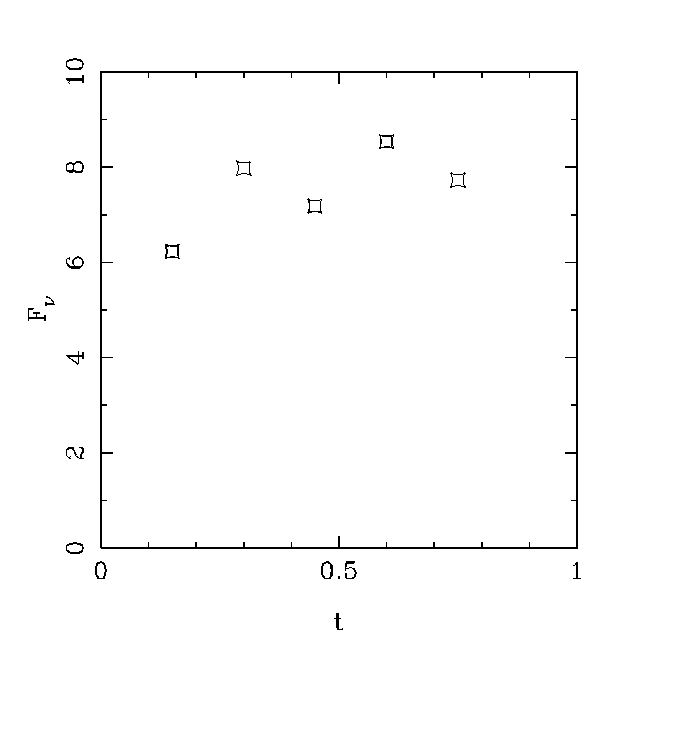}
\caption{Variability of the gamma-ray emission from the inner knot 
in computer simulations \citep{C09}. The five colour plots show 
the images of the inner knot at $\Ep=100\,$MeV ($I_\nu$ in linear scale). 
They are separated by 0.15~year, the time increasing from left to right 
and from top to bottom. 
The plot in the right bottom corner shows the corresponding 
total flux variation at $\Ep=100\,$MeV.   
}
\label{fig:var-sim}
\end{figure*}
  
The strong variability of the termination shock discovered in the numerical 
simulations by \citet{C09} and associated with the wisp production could be behind the 
observed variability of the gamma-ray emission from the Crab Nebula. 
Figure~\ref{fig:var-sim} shows the intra-year variability of the inner knot 
at $\Ep=100$~MeV based on the results of these simulations. The epoch 
corresponds to the present age of the Crab Nebula. Within 
this particular period the total flux changes were limited by $\simeq27\%$. 
(Unfortunately, most of the simulation data is now lost and we cannot comment 
on the statistical significance of this result.) The data also indicate noticeable 
changes in the knot appearance and small changes in its location.    

The mechanism of this shock variability is not very clear. 
It seems to be related to the unsteady 
axial pinch, which is behind the origin of the Crab jet in this model. As 
we have mentioned already, the magnetic field in the backflow at the base of the jet 
is highly inhomogeneous and this results in strong spatial and temporal 
fluctuations of the magnetic hoop stress, and hence the axial pressure. 
As the result, the arch shock dives towards the pulsar at times when the 
pressure is high, and moves further out when it is low. Another factor is  
the presence of strong vortices in the backflow, which can appear all the way along the 
arch-shock. In their eyes the pressure is lower and on the outside it is higher. 
Moreover, there are significant fluctuations of the ram pressure 
inside the simulated nebula as well. 

\begin{figure}
\begin{center}
 \includegraphics[width=70mm,angle=0]{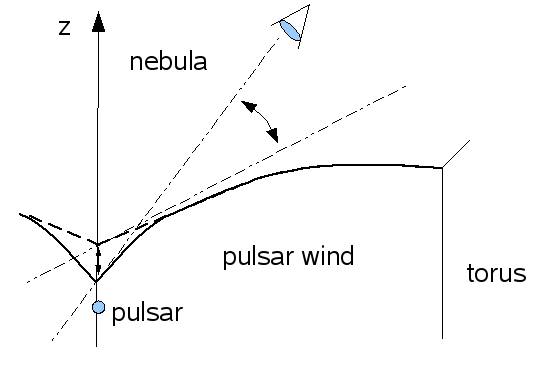}
\end{center}
\caption{ Variability of the termination shock as a reason behind 
the variability of high energy emission from the Crab Nebula. 
The solid and dotted lines show the termination shock with two extreme locations 
of its polar cusp. The dash-dotted lines show the corresponding directions of 
the Doppler beam.   
}
\label{fig:var}
\end{figure}

In the simulations, the wisps are associated with regions of high magnetic 
field in the unsteady outflow from the termination shock. New wisps are produced 
approximately on the light crossing time of the termination shock, which was 
around 10 months in the simulations and which is around 3-4 months for 
the Crab Nebula. Strong variations of the shock structure occurred on 
the similar time scale. Thus, the variability of gamma-ray emission on the time-scale 
around several months may well have this origin.  The variations of gamma-ray flux 
may be attributed to changes of the proper emissivity of the inner knot, 
associated with changes of the magnetic field strength and the number density 
of emitting particles, but also to changes in the direction of the Doppler beam 
(see Figure~\ref{fig:var}).     

Assuming the power law spectral distribution for the emissivity in the 
comoving frame, $j_\nu \propto \nu^{-\alpha}$, the observed emissivity is 
$$
   j_{\nu,\rm ob} = D^{2+\alpha} j_\nu
$$                  
\citep{LB85}.
When the angle between the flow direction and the line of sight decreases 
from $\phi=1/\gamma$ to $\phi=0$ the Doppler factor increases from
$\gamma$ to $2\gamma$. Thus, the difference in the inclination angle of the 
arch shock at the location of inner knot $\Delta\phi=1/\gamma$ can bring about the 
difference in the boosting factor up to  $2^{2+\alpha}$. 
Above $100\,$MeV, where the observed synchrotron spectrum can be approximated by  
a power law with $\alpha\sim 3$ \citep{A09}, this corresponds to a 30-fold 
flux variation. At lower photon energies, where the knot spectral index is expected to 
be $\alpha\simeq 0.6$, the corresponding flux variation is five times lower. 
Moreover, the cooling time of the electrons, producing such photons, 
becomes significantly larger --  it is already several years for the electrons 
emitting at $1\,$MeV. After traveling for such a long time 
the emitting plasma enters the remote parts of the nebula, where it inevitably 
decelerates and its emission is no longer subject to strong Doppler beaming\footnote{ 
The observed deceleration of the Crab wisps is a clear 
confirmation of such evolution \citep{H02}.}. 
The contribution of this unbeamed emission may explain the observations by 
\citet{J96}, who noticed that during these observations  the flux in the 1-30~MeV 
energy range was increasing, whereas in the 30-150~MeV range it was decreasing. 
Most likely, the total flux from the nebula at 1-30~MeV is dominated by
few recently produced wisps.

At 100~keV the observed total flux  from the Crab Nebula is 
$\simeq 4\times 10^{-28}\mbox{erg}\,\mbox{s}^{-1}\mbox{cm}^{-2}\mbox{Hz}^{-1}$ 
\citep{HA04}, whereas the expected flux from the inner knot is only 
$\simeq 1.3\times 10^{-30}\mbox{erg}\,\mbox{s}^{-1}\mbox{cm}^{-2}\mbox{Hz}^{-1}$. 
The last estimate is obtained from the power law $F_\nu \propto \nu^{-0.64}$, 
normalised using the observed optical flux of the inner knot \citep{T09}.  
Thus, even a 10-fold increase of the 
X-ray emission from the knot would produce a variation below $5\%$ in the 
total flux. 
For the similar reason, the knot variability could hardly be seen in 10~GeV-10~TeV range, 
which is dominated by the Inverse Compton emission of old electrons occupying  the  
whole volume of the nebula \citep{J96}. 


Finally, few words have to be said on the mysterious gamma-ray flares
from the Crab Nebula. The observed linear sizes of the inner knot are  
$l_\perp = \psi_\perp d_n \sim 6$ light days and $l_\parallel = \psi_\parallel 
d_n \sim 2$ light days. This implies that the variability time-scale of the 
order of few days is possible, at least in principle. 
The fact that the flare spectrum extends beyond the highest 
characteristic energy allowed for the synchrotron emission 
can still be explained by the Doppler effect.  
Fermi reported the spectral index of the flaring component $\alpha\sim 1.5$. 
Such a steep spectrum is expected because of the proximity of the 
exponential cut-off. The fact that flares are not seen at both lower and higher 
energies can be explained in the same fashion as for the long timescale variability.    

The most difficult task is to explain not only the 
short duration of the flares but also the fact that they are quite rare. 
If indeed they originated from the inner knot then this had to be associated with 
some rather peculiar events. For example, they could be produced when 
some explosive event at the base of Crab's jet drives a shock, which then collides 
with the termination shock near the knot location, with both shocks being almost 
parallel to each other.  The shocks intersection point could move towards the observer 
with speed very close to the speed of light, potentially resulting in a short burst 
of emission associated with this point.  
The magnetic reconnection could be behind such explosions but this would 
probably require a significantly higher magnetization compared to what is assumed in 
the current MHD models.               
So far, the emission from simulated PWN was computed simply by
integrating the instantaneous emissivity along the line of sight. The relativistic 
retardation effect was not taken into account. This practice has to be abandoned 
in future studies as it filters out the short time-scale variability associated with 
the relativistic motion along the line of sight.

\section{Conclusions}
\label{sec:conclusions}

\begin{enumerate} 
\item 
Downstream of oblique termination shocks of pulsar winds, the Lorentz 
factor of bulk motion can be rather high even for low-sigma winds, 
up to $\gamma=5$ for reasonably small inclination angles. This is 
shown using the analytical solution for oblique relativistic MHD shocks
and confirmed by numerical simulations of the Crab Nebula. For a high-sigma 
wind the Lorentz factor can be even higher. 
\item 
The bright inner knot in the synthetic synchrotron maps of the Crab Nebula, 
obtained in relativistic MHD simulations, is definitely a highly Doppler-boosted 
emission from the region located downstream of the termination shock (the so-called arch shock) 
and close to its polar cusp. The inner knot of the Crab Nebula is likely to be 
of the same origin.  Its geometrical parameters, such as the ratio of its major axis to 
its distance from the pulsar, and the ratio of its minor and major axes, are consistent 
with this interpretation. 
\item 
The combination of the short synchrotron cooling time of gamma-ray emitting 
electrons and the strong Doppler beaming in the vicinity of the 
termination shock suggest that if the synchrotron gamma-ray electrons are 
accelerated mainly at the termination shock then the inner knot makes 
a major contribution to the integral gamma ray emission from the Nebula around 
100~MeV.    
The two-point spectral index, $\alpha\simeq0.64$, based on  the observed optical 
emission of the knot and the integral gamma-ray emission of the Crab Nebula at 100~MeV,
is consistent with this hypothesis. A similar value is obtained in the ``standard model'' 
of the Crab Nebula emission by \citet{KC84} for the spectrum injected into the nebula 
by the termination shock.  
\item 
The observed variability of the Crab Nebula in the 1-100~MeV window on the timescale 
from one month to several years can be related to the large scale variability of the 
termination shock discovered in recent high resolution numerical simulations. 
\item 
The small size of the inner knot of the Crab Nebula, 2-6 light days, 
show that the recently discovered short gamma-ray flares can also originate
from the knot. However, the exact mechanism behind such short and rare 
events remains unclear.    
\end{enumerate}

The most critical prediction of our model, which allows a relatively 
simple test with currently available telescopes, is that the un-pulsed  
synchrotron gamma-ray emission of the Crab Nebula 
in the Fermi and AGILE windows originates from within one arcsecond of the pulsar
itself. Although the angular resolution of gamma-ray telescopes is not even 
close to one arcsecond, the test can be based on comparing the gamma-ray 
light curve with the one obtained in optics for the inner knot.
Within one arcsecond the inner knot is the dominant feature, 
apart from the pulsar itself. A potential problem of this test is the close 
proximity of the knot to the pulsar, which makes image based separation of 
their fluxes rather tricky,  even for HST and the ground-based instruments with adaptive 
optics. However, this does not seem to be needed as the pulsed emission from the Crab 
pulsar is very stable and its un-pulsed emission is not expected to be variable 
too. Hence, one only needs to measure the total flux from within $\sim 1\arcsec$ 
of the pulsar and subtract from it the phase averaged flux of the pulsed emission.   
The optical variability is expected to be strong, 
as the typical flux from the the inner knot is already about $6-10\%$ of the flux 
from the pulsar flux.  The light curves in optics and gamma rays should correlate.


\appendix

\section{Relativistic Oblique MHD Shocks}
\label{sec:obl}

The first issue is how high can be the Lorentz factor downstream 
of an oblique shock. In the limit of vanishing magnetization, oblique 
relativistic shocks was studied by \citet{K80}. The general case of relativistic 
MHD shocks was considered in \citet{MA87}, whereas \citep{Lt04} explored the 
special case where the magnetic field is parallel to the shock front. 
Here we deal only with this case, as the magnetic field of pulsar winds is almost 
perfectly asymuthal and hence parallel to the axisymmetric wind termination shock, 
and focus on the question of how relativistic can be the flow behind this shock. 

In the shock frame, the fluxes of energy, momentum, rest mass, and 
magnetic field are continuous across the shock 
\begin{equation}
\label{s1}
(w+B^2)\gamma^2\beta_x = \mbox{const},
\end{equation}
\begin{equation}
(w+B^2)\gamma^2\beta_x\beta_x +p +\frac{B^2}{2} = \mbox{const},
\label{s2}
\end{equation}
\begin{equation}
(w+B^2)\gamma^2\beta_x\beta_y = \mbox{const},
\label{s3}
\end{equation}
\begin{equation}
\rho\gamma\beta_x = \mbox{const},
\label{s4}
\end{equation}
\begin{equation}
B\gamma\beta_x = \mbox{const},
\label{s5}
\end{equation}
where $\rho$ is the rest mass density, $p$ is the gas pressure, $w=\rho c^2 + \kappa P$ is 
the relativistic enthalpy, $\kappa=\Gamma/(\Gamma-1)$, where $\Gamma$ is the adiabatic 
index, $B$ is the magnetic field as measured in the fluid frame,   
$\beta=v/c$, and $\gamma$ is the Lorentz factor. 
We select the frame where the velocity vector is in the xy-plane, the magnetic field 
is parallel to the z-direction, and the shock front is parallel to the yz-plane.  
In what follows we will use subscripts 1 and 2 to denote the upstream an the downstream 
states respectively. 

From Equations (\ref{s1}) and (\ref{s3}) it follows that 
\begin{equation}
   \beta_{1y}=\beta_{2y},
\label{os1}
\end{equation}
whereas Equations (\ref{s1}),(\ref{s4}) and (\ref{s5}) yield 
\begin{eqnarray}
\frac{\rho_2}{\rho_1}=\frac{B_2}{B_1}=\frac{\sigma_2}{\sigma_1}=\frac{\eta}{\chi},
\label{os2}
\end{eqnarray}
\begin{equation}
a_2^2=\frac{1}{\kappa} \left[  
  \eta(1+\kappa a_1^2 + \sigma_1)-\sigma_1\fracb{\eta}{\chi} -1,
\right] 
\label{os3}
\end{equation}
where $\sigma = B^2/4\pi\rho c^2$ is the magnetization parameter, 
$a^2=p/\rho c^2$ is the temperature parameter, 
$\chi=\beta_{2x}/\beta_{1x}$, and  
\begin{equation}
   \eta=\frac{\gamma_1}{\gamma_2}=(1+u_{1x}^2(1-\chi^2))^{1/2}.
\label{s6}
\end{equation}
Equations (\ref{os1}-\ref{s6}) allow to find
the downstream state given the parameters of the upstream state and $\chi$. 
The equation for $\chi$ is derived via substituting the expressions for $\rho_2$, 
$\sigma_2$, and $a_2$ from  Equations (\ref{os1}-\ref{os3}) 
into Equation (\ref{s2}), which can be written as 
\begin{eqnarray}
\nonumber
(1+\kappa a_1^2+\sigma_1)\frac{j^2}{\rho_1} + \rho_1(a_1^2+\frac{1}{2}\sigma_1) = 
\end{eqnarray}
\begin{equation} 
\qquad\qquad
(1+\kappa a_2^2+\sigma_2)\frac{j^2}{\rho_2} + \rho_2(a_2^2+\frac{1}{2}\sigma_2), 
\label{os4}
\end{equation}
where $j=\rho\gamma\beta_x$. 
In general, this is a rather combersome algebraic equation which has to be 
solved numerically. However, the pulsar winds are expected to be cold ($a_1 \to 0$) 
and highly relativistic ($\gamma_1\gg 1$), which allows significant 
simplifications. In the limit of cold upstream flow, this equation reduces to  
\begin{eqnarray}
\nonumber
\chi^3 \left[(1+\sigma_1)u_{1x}^2\fracb{\kappa-1}{\kappa}\right] +
\end{eqnarray}
\begin{eqnarray}
\nonumber
\chi^2 \left[-\sigma_1u_{1x}^2 \fracb{\kappa-2}{2\kappa} 
 -(1+\sigma_1)u_{1x}^2 -\frac{\sigma_1}{2} \right]+ 
\end{eqnarray}
\begin{eqnarray}
\nonumber
\chi^1 \left[\frac{(1+\sigma_1)(1+u_{1x}^2)}{\kappa} - \frac{\eta}{\kappa} \right]+
\end{eqnarray}
\begin{equation}
\chi^0 \left[\sigma_1 \fracb{\kappa-2}{2\kappa} (1+u_{1x}^2) \right] =0.
\label{os5}
\end{equation} 
%
             
\begin{figure}
 \includegraphics[width=77mm]{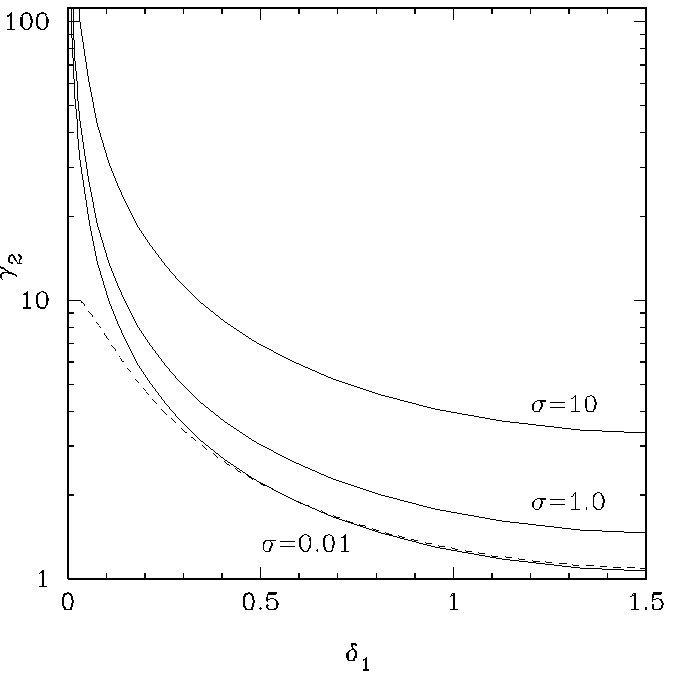}
\caption{Downstream Lorentz factor of oblique shocks as a function of the 
shock angle. Solid lines correspond to the upstream Lorentz factor   
$\gamma_1=10^6$ and the magnetization parameter 
$\sigma=0.01$, 1.0, and 10. The dashed line corresponds to $\gamma_1=10$ 
and $\sigma=0.01$.
}    
\label{fig:os}
\end{figure}
             
Further simplification can be made if $u_{1x}\gg 1$. If we denote the angle 
between the shock plane and the velocity vector as $\delta$, then this 
condition implies $\delta_1\gg 1/\gamma_1$.  In this case, the above equation
reduces to the cubic equation for $\chi$ 
\begin{eqnarray}
\nonumber
\chi^3 (1+\sigma_1)(\kappa-1) - 
\chi^2 \left(\sigma_1\frac{3\kappa-2}{2} +\kappa \right)+
\end{eqnarray}
\begin{equation}
\qquad\qquad
\chi (1+\sigma_1) + \sigma_1 \fracb{\kappa-2}{2}=0
\label{os6}
\end{equation}
Its root $\chi=1$ corresponds to the continious solution. 
Of the other two roots only 
\begin{equation}
\chi=\frac{2+6\sigma_1 + \sqrt{\cal D}}{4(1+\sigma_1)(\kappa-1)},
\label{os7}
\end{equation}
where
$$
{\cal D} = (2+\sigma_1 \kappa)^2 - 8(1+\sigma_1)\sigma_1(\kappa-1)(2-\kappa),  
$$
is physical. 
For the ultrarelativistic equation of state, with $\Gamma=4/3$ and $\kappa=4$, 
this becomes 
\begin{equation}
\chi=\frac{1+2\sigma_1 + \sqrt{16\sigma_1^2+16\sigma_1+1}}{6(1+\sigma_1)}.
\label{os8}
\end{equation}
The corresponding downstream Lorentz factor is 
\begin{equation}
   \gamma_2  = \frac{1}{\sqrt{1-\chi^2}} \frac{1}{\sin\delta_1}.   
\label{os9}
\end{equation}
For $\sigma_1\gg 1$ this yields 
\begin{equation}
   \gamma_2  \simeq \frac{\sigma_1^{1/2}}{\sin\delta_1},
\label{os10}
\end{equation}
and for $\sigma_1\ll 1$ 
\begin{equation}
   \gamma_2  \simeq \frac{3}{2\sqrt{2}} (1+\frac{1}{2}\sigma_1) 
   \frac{1}{\sin\delta_1}.
\label{os11}
\end{equation}
Figure~\ref{fig:os} shows $\gamma_2(\delta_1)$ for $\sigma_1=0,1$ and 10. 

The downstream shock angle 
\begin{equation} 
   \tan\delta_2  =  \chi(\sigma_1)\tan\delta_1. 
\label{os12}
\end{equation}
For $\sigma\ll 1$ and in the of small angle approximation this 
yields 

\begin{equation}
   \delta_2  \simeq  \frac{1}{3} \delta_1.
\label{os13}
\end{equation}

The general conlusion from this analysis is that the downstream flow 
can be highly relativistic, provided the shock is sufficiently oblique. 
The current two-dimensional MHD models of PWN do predict a very oblique termination 
shock due the anisotropic distribution of energy flux in the pulsar wind. For $\sigma_1\ll 1$ 
in the pulsar wind, the case favoured by the models, and in the small angle 
approximation, we have $\gamma_2\simeq 1/\delta_1$. The Lorentz factor can be significantly 
higher if $\sigma>1$.  
 
In the numerical simulations cited above the Lorentz factor of pulsar wind 
was rather low, $\gamma_1\simeq10$. The results for such a relatively low 
upstream Lorentz factor are shown in Figure~\ref{fig:os} by the dashed line.  
One can see, that this really becomes a factor only for very small shock 
angles, $\delta_1<0.2$.

\end{document}